# Energetic Electron Irradiations of Amorphous and Crystalline Sulphur-Bearing Astrochemical Ices

Duncan V. Mifsud[1,2,*], Péter Herczku[2,*], Richárd Rácz[2], K.K. Rahul[2], Sándor T.S. Kovács[2], Zoltán Juhász[2], Béla Sulik[2], Sándor Biri[2], Robert W. McCullough[3], Zuzana Kaňuchová[4], Sergio Ioppolo[5], Perry A. Hailey[1], and Nigel J. Mason[1,2,*]

1. *Centre for Astrophysics and Planetary Science, School of Physical Sciences, University of Kent, Canterbury CT2 7NH, United Kingdom*

2. *Institute for Nuclear Research (Atomki), Debrecen H-4026, Hungary*

3. *Department of Physics and Astronomy, School of Mathematics and Physics, Queen's University Belfast, Belfast BT7 1NN, United Kingdom*

4. *Astronomical Institute, Slovak Academy of Sciences, Tatranská Lomnica SK-059 60, Slovakia*

5. *School of Electronic Engineering and Computer Science, Queen Mary University of London, London E1 4NS, United Kingdom*

\*    Correspondence:   Duncan V. Mifsud    dm618@kent.ac.uk
                           Péter Herczku    herczku@atomki.hu
                           Nigel J. Mason    n.j.mason@kent.ac.uk

### ORCID Identification Numbers

| | |
|---|---|
| Duncan V. Mifsud | 0000-0002-0379-354X |
| Péter Herczku | 0000-0002-1046-1375 |
| Richard Rácz | 0000-0003-2938-7483 |
| K.K. Rahul | 0000-0002-5914-7061 |
| Sándor T.S. Kovács | 0000-0001-5332-3901 |
| Zoltán Juhász | 0000-0003-3612-0437 |
| Béla Sulik | 0000-0001-8088-5766 |
| Sándor Biri | 0000-0002-2609-9729 |
| Robert W. McCullough | 0000-0002-4361-8201 |
| Zuzana Kaňuchová | 0000-0001-8845-6202 |
| Sergio Ioppolo | 0000-0002-2271-1781 |
| Perry A. Hailey | 0000-0002-8121-9674 |
| Nigel J. Mason | 0000-0002-4468-8324 |

**Abstract**

Laboratory experiments have confirmed that the radiolytic decay rate of astrochemical ice analogues is dependent upon the solid phase of the target ice, with some crystalline molecular ices being more radio-resistant than their amorphous counterparts. The degree of radio-resistance exhibited by crystalline ice phases is dependent upon the nature, strength, and extent of the intermolecular interactions that characterise their solid structure. For example, it has been shown that crystalline $CH_3OH$ decays at a significantly slower rate when irradiated by 2 keV electrons at 20 K than does the amorphous phase due to the stabilising effect imparted by the presence of an extensive array of strong hydrogen bonds. These results have important consequences for the astrochemistry of interstellar ices and outer Solar System bodies, as they imply that the chemical products arising from the irradiation of amorphous ices (which may include prebiotic molecules relevant to biology) should be more abundant than those arising from similar irradiations of crystalline phases. In this present study, we have extended our work on this subject by performing comparative energetic electron irradiations of the amorphous and crystalline phases of the sulphur-bearing molecules $H_2S$ and $SO_2$ at 20 K. We have found evidence for phase-dependent chemistry in both these species, with the radiation-induced exponential decay of amorphous $H_2S$ being more rapid than that of the crystalline phase, similar to the effect that has been previously observed for $CH_3OH$. For $SO_2$, two fluence regimes are apparent: a low-fluence regime in which the crystalline ice exhibits a rapid exponential decay while the amorphous ice possibly resists decay, and a high-fluence regime in which both phases undergo slow exponential-like decays. We have discussed our results in the contexts of interstellar and Solar System ice astrochemistry and the formation of sulphur allotropes and residues in these settings.

# 1    Introduction

It has been established for some time now that the laboratory irradiation of astrochemical ice analogues using energetic charged particles (i.e., ions and electrons) or ultraviolet photons may lead to the production of prebiotic molecules relevant to biology, such as amino acids or nucleobases (e.g., Muñoz-Caro *et al.* 2002, Hudson *et al.* 2008, Nuevo *et al.* 2012). Motivated by a desire to further understand the non-equilibrium chemistry leading to the formation of these so-called 'seeds of life', many studies have sought to determine and quantify the influence of various physical parameters on the outcome of such reactions. Perhaps the best studied of these is ice temperature, with previous works having demonstrated the key influence of this parameter on the abundance of product molecules formed after irradiation (e.g., Sivaraman *et al.* 2007, Mifsud *et al.* 2022a).

Our recent work has also demonstrated that the solid phase of an irradiated ice plays a crucial role in determining the outcome of astrochemical reactions mediated by ionising radiation. Through a series of comparative electron irradiations, we have demonstrated that the radiolytic decay rate of an astrochemical ice is dependent upon the nature, strength, and extent of the intermolecular interactions that characterise its solid phase (Mifsud *et al.* 2022b, Mifsud *et al.* 2022c). For instance, the decay rate of α-crystalline $CH_3OH$ was found to be significantly less rapid than that of the amorphous phase. This was attributed to the existence of an extensive network of strong hydrogen bonds that exists in the α-crystalline phase. This network requires an additional energy input from the projectile electrons to be overcome, thus leaving less energy overall to drive radiolytic chemistry. Conversely, the amorphous $CH_3OH$ ice is characterised only by localised clusters of hydrogen bonded molecules. Such a structure does not benefit from the same stabilising effect supplied by the network of hydrogen bonds in the α-crystalline phase, particularly as hydrogen bonding in $CH_3OH$ is known to be a cooperative phenomenon in which the presence of one hydrogen bond in the network strengthens successive hydrogen bonds through electrostatic effects (Kleeberg and Luck 1989, Sum and Sandler 2000).

In the case of $N_2O$ ice, the decay rate of the amorphous phase was noted to be only moderately more rapid than that of the crystalline phase (Mifsud *et al.* 2022b). The dominant intermolecular forces of attraction in solid $N_2O$ are expected to be dipole-dipole interactions. Although the orientation of these dipoles in the crystalline phase is anticipated to confer some degree of resistance against radiolytic decay compared to the amorphous phase, this is considerably less than that induced by the hydrogen bonding network in α-crystalline $CH_3OH$. This therefore explains the more similar radiolytic decay rates of amorphous and crystalline $N_2O$. Such results carry important implications for the radiation processing of astrochemical ices, as they suggest that the irradiation of amorphous ices is more chemically productive than that of crystalline ones; particularly in the case of those ices which are able to form strong and extensive intermolecular bonds when crystalline. Extending this idea further, it is entirely possible that those astrophysical environments in which space radiation-induced amorphisation processes dominate over thermally-induced crystallisation may be characterised by a more productive radiation chemistry. This idea is not unreasonable, particularly in light of the discovery of several complex organic molecules in pre-stellar cores (e.g., McGuire *et al.* 2020, Burkhardt *et al.* 2021).

In this present study, we have expounded upon our previous work by performing comparative electron irradiations of the crystalline and amorphous phases of pure $H_2S$ and $SO_2$ astrochemical ice analogues, thus simulating the processing such ices undergo during their interaction with galactic cosmic rays, stellar winds, or magnetospheric plasmas as a result of the production of large quantities of secondary electrons (Mason *et al.* 2014, Boyer *et al.* 2016). Solid $H_2S$ is known to exhibit a number of stable crystalline phases under low temperature and ambient pressure conditions (Fathe *et al.* 2006), but it is the crystalline phase III (hereafter simply referred to as the crystalline $H_2S$ phase) which is of importance under conditions relevant to astrochemistry. This phase is orthorhombic, having eight molecules per unit cell and adopting the *Pbcm* space group. $SO_2$ may also exist as an orthorhombic crystalline solid under astrochemical conditions, but in this case the *Aba*2 space group is adopted and there are only two molecules per unit cell (Schriver-Mazzuoli *et al.* 2003).

Although sulphur is one of the most abundant elements in the cosmos and is of importance in both biochemical and geochemical contexts, much remains unknown regarding its chemistry in interstellar and outer Solar System settings (Mifsud *et al.* 2021a). It is thought, for instance, that $H_2S$ ice processing by galactic cosmic rays or ultraviolet photons accounts for the apparent depletion of sulphur (relative to its total cosmic abundance) in dense interstellar clouds by producing large quantities of atomic sulphur or molecular sulphur chains and rings which are difficult to detect using current observation techniques (Jiménez-Escobar and Muñoz-Caro 2011, Jiménez-Escobar *et al.* 2014). $H_2S$ itself has not yet been definitively detected in interstellar icy grain mantles (Boogert *et al.* 2015). Conversely, $SO_2$ ice has been detected within both the dense interstellar medium as well as on the surfaces of outer Solar System bodies such as the Galilean moons of Jupiter (Boogert *et al.* 1997, Carlson *et al.* 1999). However, the exact chemical mechanisms leading to its formation in these settings remain widely debated (Mifsud *et al.* 2021a).

The purpose of this study is thus two-fold: (i) to determine whether the phase of irradiated sulphur-bearing molecular ices influences the radiation-induced rate of decay as was previously demonstrated for non-sulphur-bearing ices; and (ii) to contribute further to our (comparatively poor) understanding of the extra-terrestrial chemistry of sulphur. To achieve these goals, the amorphous and crystalline phases of pure $H_2S$ and $SO_2$ ices were respectively irradiated with 2 and 1.5 keV electrons, and the resultant physico-chemical changes were followed *in situ* using Fourier-transform mid-infrared (FT-IR) transmission absorption spectroscopy.

## 2  Experimental Methodology

The irradiation experiments were performed using the Ice Chamber for Astrophysics-Astrochemistry (ICA); a custom-built experimental apparatus located at the Institute for Nuclear Research (Atomki) in Debrecen, Hungary. This apparatus (Fig. 1) has been described in detail in previous publications (Herczku *et al.* 2021, Mifsud *et al.* 2021b), and so only a brief description of the most salient details will be provided here. The ICA is a UHV-compatible chamber with a nominal base pressure of a few $10^{-9}$ mbar which is achieved by the combined action of a dry rough vacuum pump and a turbomolecular pump. Within the centre of the chamber is a gold-coated oxygen-free copper sample holder which supports up to four ZnSe deposition substrates, onto which astrochemical ice analogues may be prepared. The temperature of the sample holder and the substrates may be cooled to 20 K using a closed-cycle helium cryostat, although an operational temperature range of 20-300 K is available.

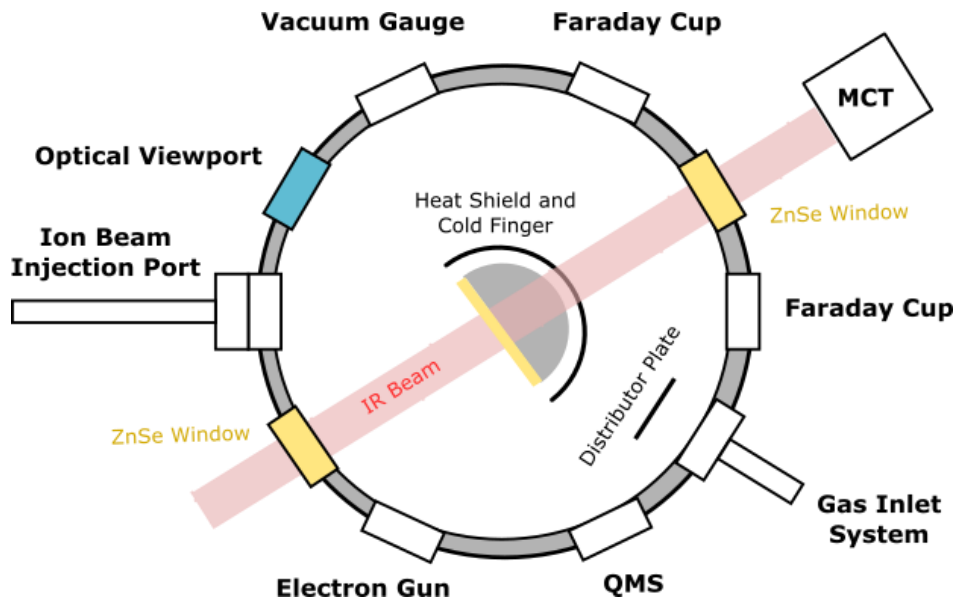

**Fig. 1:** Top-view schematic diagram of the ICA set-up. Note that electron irradiations are carried out such that projectile electrons impact the target ices at 36° to the normal. Figure reproduced from Mifsud *et al.* (2021b) with the kind permission of the European Physical Journal (EPJ).

The preparation of $H_2S$ and $SO_2$ astrochemical ice analogue phases was achieved *via* background deposition by allowing the relevant gas into a pre-mixing line before dosing it into the main chamber at a pressure of a few $10^{-6}$ mbar. Amorphous ice phases were prepared by deposition at 20 K, while crystalline $H_2S$ and $SO_2$ ices were prepared by deposition at 60 and 90 K, respectively, before being cooled to 20 K. Once deposited, FT-IR spectra (spectral range = 4000-650 $cm^{-1}$; spectral resolution = 1 $cm^{-1}$) of the ices were acquired, from which quantitative measurements of their molecular column densities $N$ (molecules $cm^{-2}$) and thicknesses $d$ (μm) could be performed by measuring the peak area $P$ ($cm^{-1}$) of a characteristic absorption band (Eq. 1):

$$d = 10{,}000 \times \frac{NZ}{N_A \rho} = 10{,}000 \times \ln(10) \times \frac{PZ}{A_v N_A \rho}$$

(Eq. 1)

where $Z$ is the molar mass (g $mol^{-1}$) of the molecular ice, $N_A$ is the Avogadro constant ($6.02 \times 10^{23}$ molecules $mol^{-1}$), $\rho$ is the density of the ice (g $cm^{-3}$), and $A_v$ is the band strength constant of the characteristic absorption band whose area is being measured (cm $molecule^{-1}$). Information on the molecular column densities and thicknesses of the ices investigated in this study, as well as the physical parameters used to calculate these values, is given in Tables 1 and 2.

The deposited pure $H_2S$ and $SO_2$ astrochemical ices were respectively irradiated using 2 and 1.5 keV electron beams (average flux = $4 \times 10^{13}$ electrons $cm^{-2}$ $s^{-1}$) to a total fluence of about $8.3 \times 10^{16}$ electrons $cm^{-2}$, with projectile electrons impacting the target ices at an angle of 36° to the normal. Prior to commencing the irradiations, the beam current, spot size, and profile homogeneity were determined using the method described by Mifsud *et al.* (2021b). CASINO simulations (Drouin *et al.* 2007) of the trajectory of the electrons as they travelled through the solid ices revealed that the maximum penetration depths of the incident electrons into the $H_2S$ and $SO_2$ ices were 155 and 70 nm, respectively. FT-IR spectra were collected at several intervals throughout the irradiation process so as to monitor the radiation chemistry occurring. All irradiations were carried out at 20 K so as to preclude any temperature-dependent effects on the mobility of radiolytically derived radicals. Moreover, the irradiation of each ice phase was performed three times so as to ensure good repeatability of the experiment.

**Table 1:** List of physical parameters and constants used for the quantitative study of the deposited $H_2S$ and $SO_2$ astrochemical ices.

| Physical Parameter | $H_2S$ | $SO_2$ | Reference |
|---|---|---|---|
| Absorption Band Position (cm$^{-1}$) | 2550 | 1148 | Garozzo *et al.* (2008) and Hudson and Gerakines (2018) |
| Amorphous $A_v$ (10$^{-17}$ cm molecule$^{-1}$) | 1.12 | 0.22 | Garozzo *et al.* (2008) and Hudson and Gerakines (2018) |
| Crystalline $A_v$ (10$^{-17}$ cm molecule$^{-1}$) | 2.90 | 0.88 | Garozzo *et al.* (2008) and Hudson and Gerakines (2018) |
| Amorphous $T_{\text{deposition}}$ (K) | 20 | 20 | This work |
| Crystalline $T_{\text{deposition}}$ (K) | 60 | 90 | This work |
| $T_{\text{irradiation}}$ (K) | 20 | 20 | This work |
| $Z$ (g mol$^{-1}$) | 34 | 64 | This work |
| Density (g cm$^{-3}$) | 1.22 | 1.89 | Post *et al.* (1952) and Yarnall and Hudson (2022) |
| $E_{\text{electron}}$ (keV) | 2.0 | 1.5 | This work |
| Maximum Electron Penetration Depth (nm) | 155 | 70 | Drouin *et al.* (2007) |

**Table 2:** List of initial molecular column densities and thicknesses of the $H_2S$ and $SO_2$ ices investigated in this study.

| Ice | Species | Phase | $N$ (10$^{17}$ molecules cm$^{-2}$) | $d$ (μm) |
|---|---|---|---|---|
| 1 | $H_2S$ | Amorphous | 7.05 | 0.326 |
| 2 | $H_2S$ | Amorphous | 6.50 | 0.301 |
| 3 | $H_2S$ | Amorphous | 7.67 | 0.355 |
| *Average* | | | *7.07* | *0.327* |
| 4 | $H_2S$ | Crystalline | 5.78 | 0.268 |
| 5 | $H_2S$ | Crystalline | 6.35 | 0.294 |
| 6 | $H_2S$ | Crystalline | 7.61 | 0.352 |
| *Average* | | | *6.58* | *0.305* |
| 7 | $SO_2$ | Amorphous | 3.09 | 0.174 |
| 8 | $SO_2$ | Amorphous | 2.49 | 0.140 |
| 9 | $SO_2$ | Amorphous | 2.89 | 0.162 |
| *Average* | | | *2.82* | *0.159* |
| 10 | $SO_2$ | Crystalline | 2.62 | 0.147 |
| 11 | $SO_2$ | Crystalline | 2.19 | 0.123 |
| 12 | $SO_2$ | Crystalline | 2.84 | 0.160 |
| *Average* | | | *2.55* | *0.143* |

## 3 Results and Discussion

The FT-IR spectra of the pure $H_2S$ and $SO_2$ ice phases investigated in this study, both before and after irradiation by electrons at different fluences, are depicted in Fig. 2. In the amorphous phase, $H_2S$ presents a very broad absorption band which peaks at 2550 cm$^{-1}$ attributable to both the symmetric ($v_1$) and asymmetric ($v_3$) stretching modes. In the crystalline phase, this band is better resolved and the individual contributors may be observed. The asymmetric stretching mode is observed to peak at 2546 cm$^{-1}$, while the symmetric stretching mode is split into two components peaking at 2534 and 2522 cm$^{-1}$. Fathe et al. (2006) also observed this splitting and attributed it to the existence of two unique sulphur atoms and three unique S–H bonds in the unit cell.

The amorphous phase of solid $SO_2$ presents two distinct albeit broad absorption asymmetric and symmetric stretching mode bands which respectively peak at 1320 and 1148 cm$^{-1}$. In the crystalline phase, these bands are observed to be better resolved, with three and two individual structures being observed in the asymmetric and symmetric stretching mode bands, respectively. These structures have been attributed to the various isotopologues of $SO_2$ (Schriver-Mazzuoli et al. 2003): the bands peaking at 1323, 1309, 1303, and the shoulder band at 1301 cm$^{-1}$ are ascribed to the transverse $B_1$(TO) and $B_2$(TO) modes of $^{32}S^{16}O_2$, and to $^{34}S^{16}O_2$ and $^{32}S^{16}O^{18}O$; while those peaking at 1143 and 1140 cm$^{-1}$ are respectively attributed to the transverse $A_1$(TO) mode of $^{32}S^{16}O_2$ and to $^{34}S^{16}O_2$. It is interesting to note that the naturally low abundances of $^{34}S$ and $^{18}O$ do not result in weaker band intensities for the $SO_2$ isotopologues containing these isotopes due to intermolecular coupling between these isotopologues in the condensed phase (Brooker and Chen 1991, Schriver-Mazzuoli et al. 2003).

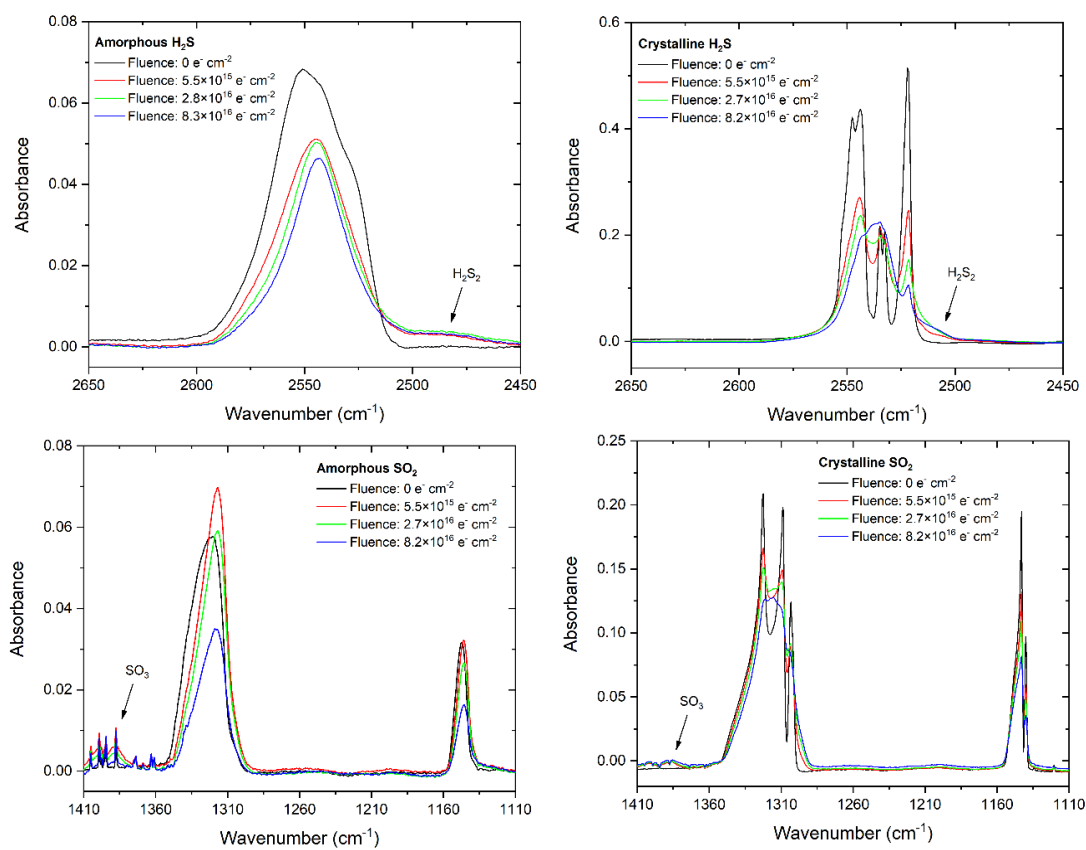

**Fig. 2:** FT-IR spectra of the amorphous and crystalline phases of $H_2S$ and $SO_2$ ices at several points during their irradiation by energetic electrons at 20 K. Note that the fine structures coincident with the $SO_3$ absorption band in the spectrum of the electron irradiated amorphous $SO_2$ ice are caused by instabilities in the purge of the detector. Moreover, the initial increase in the intensity of the amorphous $SO_2$ asymmetric stretching mode is likely caused by the radiation-induced compaction of the porous ice.

The onset of electron irradiation brings about noticeable changes in the appearances of the spectra of the pristine ices. Perhaps the most prominent of these is the significant broadening of the crystalline ice absorption bands, which also lose their resolved individual structures. This is due to radiation-induced amorphisation, which has been well documented in several ice species irradiated by ions, electrons, and ultraviolet photons; including $H_2O$, $CH_3OH$, $N_2O$, and $NH_3$ (e.g., Kouchi and Kuroda 1990, Moore *et al.* 2007a, Famá *et al.* 2010, Mifsud *et al.* 2022b, Mifsud *et al.* 2022c). It is interesting to note that, even at the end of the irradiation process once a fluence of $>8\times10^{16}$ electrons cm$^{-2}$ has been delivered to the crystalline ices, the appearances of their absorption bands are still not identical to those of the deposited amorphous ices. Indeed, small signs of crystallinity (e.g., the presence of shoulder bands or shifted band peak positions) are still observable in the crystalline ices at the end of irradiation. As such, these irradiated ices are likely largely amorphous but with some small degree of remnant structural order.

The irradiation of molecular ices is known to initiate a rich chemistry leading to the formation of new species. Previous studies have established that irradiated $H_2S$ ices efficiently yield $H_2S_2$ as well as higher order polysulphanes ($H_2S_x$, where $x > 2$) in addition to allotropic forms of elemental sulphur (Shingledecker *et al.* 2020, Cazaux *et al.* 2022). In our experiments, we have observed the formation of $H_2S_2$ through the development of its vibrational stretching modes which appear as a broad shoulder band on the lower wavenumber end of the analogous $H_2S$ absorption bands at about 2500 cm$^{-1}$ (Fig. 2; Moore *et al.* 2007b). The chemistry leading to the formation of $H_2S_2$ (as well as higher order polysulphanes) is thought to be largely mediated by HS radicals formed *via* the dissociation of the parent $H_2S$ molecules:

$$H_2S \rightarrow HS + H$$

(Eq. 2)

$$2\ HS \rightarrow H_2S_2$$

(Eq. 3)

It should be noted that HS radicals produced as a result of the radiolytic dissociation of $H_2S$ may pick up an electron to form HS$^-$ ions. These HS$^-$ ions may possibly participate in chemistry leading to the formation of other HS radicals *via* proton abstraction reactions with $H_2S$, after which the resultant HS$^-$ ion may undergo electron auto-detachment to yield HS. A similar process has recently been demonstrated to occur in $H_2O$ ice with respect to radiolytically derived OH radicals and OH$^-$ ions (Kitajima *et al.* 2021).

The irradiation of the $SO_2$ ice phases was also observed to lead to the formation of new molecules; in particular $SO_3$ which was observed through its asymmetric stretching absorption band at 1388 cm$^{-1}$ (Guldan *et al.* 1995). $SO_3$ formation in irradiated $SO_2$ ices has been studied extensively and is believed to be the result of the dissociation of the latter species to yield free oxygen atoms which may then bond with other $SO_2$ molecules (Moore *et al.* 2007b). It should be noted, however, that earlier studies by Pilling and Bergantini (2015) and de Souza Bonfim *et al.* (2017) have demonstrated that electronically excited $SO_2$ may also abstract oxygen atoms from either an adjacent $SO_2$ molecule or from a $O_2$ molecule; the latter having likely been formed as a result of the double ionisation of the $SO_2$ parent molecule followed by electron neutralisation as described recently by Wallner *et al.* (2022):

$$SO_2 \rightarrow SO + O$$

(Eq. 4)

$$SO_2 + O \rightarrow SO_3$$

(Eq. 5)

$$SO_2^* + SO_2 \text{ (or } O_2\text{)} \rightarrow SO_3 + SO \text{ (or O)}$$

(Eq. 6)

Differences in the parent molecule decay trends and in the abundance of molecular products observed after irradiation were noted between the studied amorphous and crystalline ice phases. Considering first the decay trends of the amorphous and crystalline $H_2S$ ices: it was noted that the rate of decay of the crystalline phase was significantly slower than that of the amorphous phase (Fig. 3). A similar trend was observed during the comparative electron irradiations of the amorphous and crystalline phases of $CH_3OH$, $N_2O$, and $H_2O$ ices (Mifsud *et al.* 2022b, Mifsud *et al.* 2022c). This was attributed to the additional energy input required to disrupt the extensive intermolecular forces of attraction that characterise the crystalline solid before radiolytic chemistry as a result of molecular dissociation may proceed. In $CH_3OH$, the α-crystalline phase contains extensive arrays of cooperative and strong hydrogen bonds which stabilise the ice considerably against radiolytic decay compared to the amorphous solid, which is only characterised by localised hydrogen bonds (Kleeberg and Luck 1989, Sum and Sandler 2000, Mifsud *et al.* 2022b).

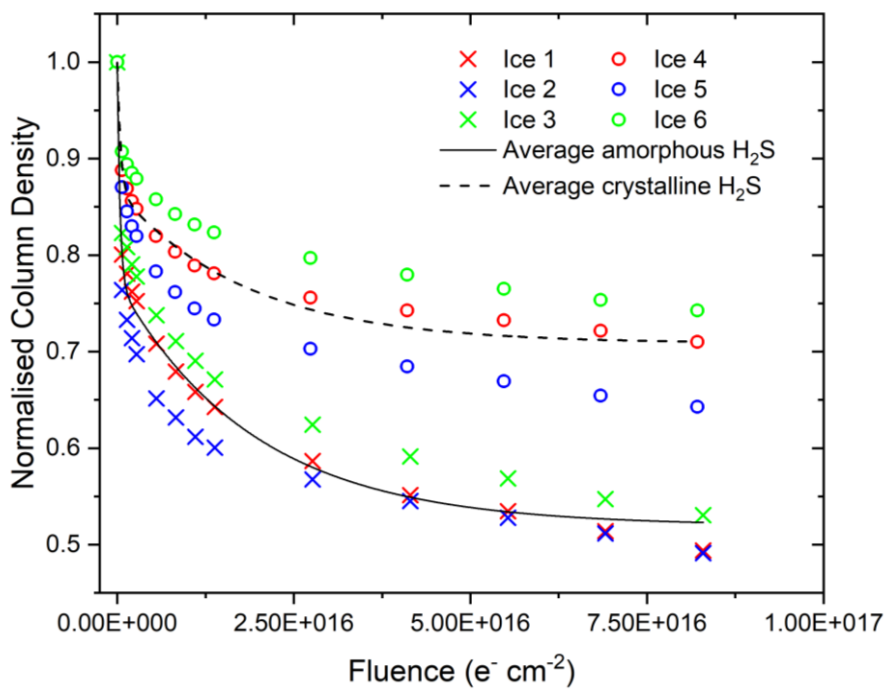

**Fig. 3:** Decay of amorphous and crystalline $H_2S$ column densities normalised to the initially deposited column density during irradiation using 2 keV electrons. Note that the average decay trends are fitted by two exponential decay functions joined at a fluence of $1.4 \times 10^{15}$ electrons cm$^{-2}$.

$H_2S$ is also capable of forming hydrogen bonds between adjacent molecules (Das *et al.* 2018), although these are significantly weaker than those in alcohols: the hydrogen bond strengths in pure $CH_3OH$ and $H_2S$ are 6.3 and 1.0 kcal mol$^{-1}$, respectively (Pellegrini *et al.* 1973, Bhattacherjee *et al.* 2013). Despite this weaker nature of the hydrogen bond in $H_2S$, it still displays a relative stability of the crystalline phase to radiation-induced decay (compared to the amorphous phase) that is qualitatively similar to that of $CH_3OH$. As such, it is likely that another factor should be invoked to account for the relative radio-resistance of the crystalline phases of these ices; that of lattice energies. The energetic advantage induced by the ordering of the molecular components of the solid ice must also be overcome and thus a proportion of the incident electrons' kinetic energy must be expended upon overcoming both the lattice energy as well as the more extensive hydrogen bonding network in the crystalline phase, leaving less energy to induce the molecular dissociation that drives radiolytic chemistry.

Such an interpretation is wholly consistent with our previously reported results on the comparative electron irradiations of amorphous and crystalline $N_2O$, which demonstrated only a moderately more rapid decay rate of the former compared to the latter (Mifsud *et al.* 2022b). In that case, a fraction of the kinetic energy of the incident electrons must be used to overcome both the increased ordering of the molecular dipoles as well as the crystal lattice energy. We also note that $A_v$ differs by a factor-of-two-and-a-half between the amorphous and crystalline $H_2S$ phases, with that of the latter being greater. This is not insignificant, and the rapid amorphisation of the crystalline phase as a result of its irradiation may mean that column density measurements of this phase may be somewhat underestimated and, as such, the radio-resistance of the crystalline phase may be even greater than that depicted in Fig. 3, although this is difficult to quantify.

The radiation-induced decay trends of amorphous and crystalline $SO_2$ (Fig. 4), however, are significantly different to those of $H_2S$ and the previously studied ices. The decay trend of the crystalline $SO_2$ ice initially exhibits the anticipated profile of a rapid exponential decay. However, once a fluence of about $1.4 \times 10^{16}$ electrons cm$^{-2}$ is exceeded, the normalised column density declines significantly more slowly. Perhaps even more surprising is the fact that the amorphous $SO_2$ normalised column density (with respect to the initial $SO_2$ column density deposited) does not really vary at low electron fluences, having an average normalised column density of 0.97 after a fluence of $8.2 \times 10^{15}$ electrons cm$^{-2}$ had been delivered. For comparison, by the point this fluence had been delivered to the crystalline $SO_2$ ice, its average normalised column density had decreased to 0.83. However, similarly to the case of the crystalline $SO_2$ ice, once a fluence of about $1.4 \times 10^{16}$ electrons cm$^{-2}$ had been delivered, the normalised column density was observed to undergo a slow exponential-like decay. Interestingly, beyond a delivered fluence of $1.4 \times 10^{16}$ electrons cm$^{-2}$, the rate of decay of the amorphous $SO_2$ is greater than that of the crystalline $SO_2$, and indeed the average decay trends cross one another at a fluence of about $5.3 \times 10^{16}$ electrons cm$^{-2}$.

Providing an exact reason for the observed amorphous $SO_2$ decay trends is a challenging task. Measurements of the photo-desorption of $SO_2$ molecules from an amorphous ice induced by soft x-rays allowed de Souza Bonfim *et al.* (2017) to suggest that, at low fluences, the recombination of fragments produced by the dissociation of $SO_2$ to yield electronically excited $SO_2$ may be a favourable process, thus largely precluding net $SO_2$ dissociation within the ice. It is also possible that the irradiation of the amorphous $SO_2$ ice results in its compaction, which may cause an increase in $A_v$ of the measured band (Fig. 2). Similar results were recently shown for ion irradiated amorphous CO ice, for which $A_v$ very rapidly increased by about 5% of its nominal value as a result of the compaction of the ice (Ivlev *et al.* 2022). It is not possible to discount either of these possible explanations based on the available evidence.

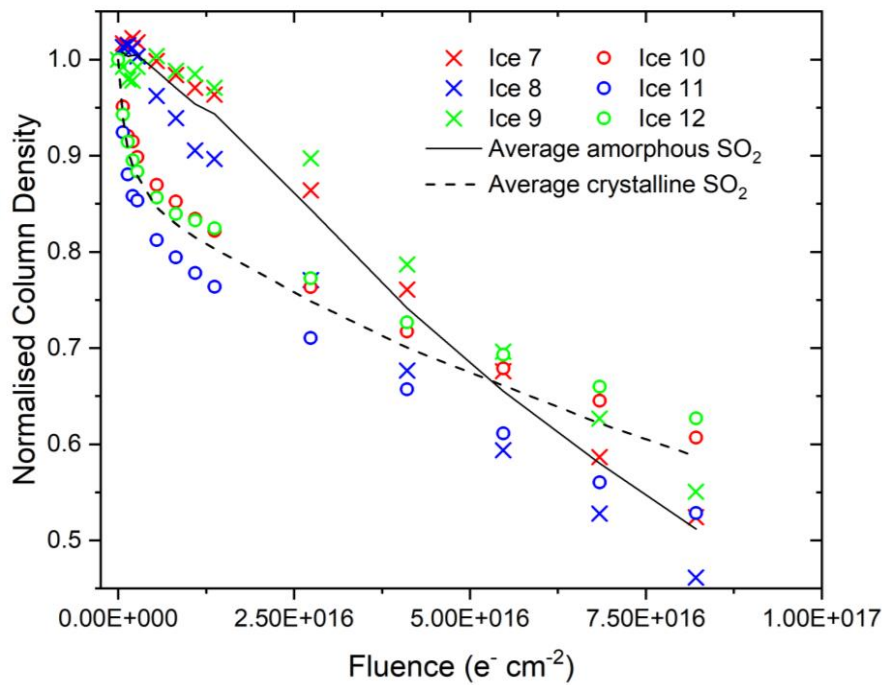

**Fig. 4:** Decay of amorphous and crystalline $SO_2$ column densities normalised to the initially deposited column density during irradiation using 1.5 keV electrons. Note that the average decay trends are not fits and are plotted solely to guide the eye.

As a final analytical consideration, we have attempted to establish the sulphur budget of the electron irradiation processes presented in this study. The possible chemical transformations of $H_2S$ and $SO_2$ to infrared inactive atomic or allotropic forms of sulphur have already been referred to (Shingledecker *et al.* 2020, Cazaux *et al.* 2022), and so it is useful to quantify how much of the initially deposited $H_2S$ or $SO_2$ ice ends up in such a form as a result of its irradiation. As depicted in Fig. 2, the only major infrared active products of $H_2S$ and $SO_2$ irradiation were $H_2S_2$ and $SO_3$, respectively. We have quantified the column densities of these product molecules throughout the irradiation processes by measuring the peak areas of their primary absorption bands and making use of Eq. 1 (Fig. 5). We note that we have taken $A_v$ for the $H_2S_2$ absorption band at about 2500 cm$^{-1}$ to be $2.4\times10^{-17}$ cm molecule$^{-1}$ (Cazaux *et al.* 2022). To the best of our knowledge, $A_v$ has not yet been defined for the $SO_3$ absorption band at 1388 cm$^{-1}$, and so we have followed the example of de Souza Bonfim *et al.* (2017) who assumed that this band strength is equal to that of the $SO_2$ asymmetric stretching mode which is $1.47\times10^{-17}$ cm molecule$^{-1}$ (Garozzo *et al.* 2008).

As expected, the yield of $H_2S_2$ from the irradiated amorphous $H_2S$ ice is greater than that from the irradiated crystalline $H_2S$ ice, commensurate with the increased decay rate of the former compared to the latter. Conversely, the electron irradiation of the crystalline $SO_2$ ice proved to be more conducive to the formation of $SO_3$ than did the irradiation of the amorphous phase. This is as expected for the low-fluence regime of the irradiation process (up to a fluence of about $5.3\times10^{16}$ electrons cm$^{-2}$), due to the amorphous $SO_2$ ice possibly resisting radiolytic decay. However, the greater abundance of $SO_3$ in the irradiated crystalline ices persists even beyond this fluence, despite the more rapid decay of amorphous $SO_2$ after this point. It should be noted, however, that after peaking at a fluence of about $5.5 \times 10^{15}$ electrons cm$^{-2}$, the $SO_3$ column density within the irradiated amorphous $SO_2$ ice also declines slightly (Fig. 5). The concomitant loss of $SO_2$ and $SO_3$ from the ice during its irradiation suggests that sulphur is either being converted into a form which is not infrared active (e.g., atomic or allotropic sulphur) or is being desorbed or sputtered from the bulk ice. In either case, however, there is a fraction of the initially deposited sulphur that remains unobserved in the ice.

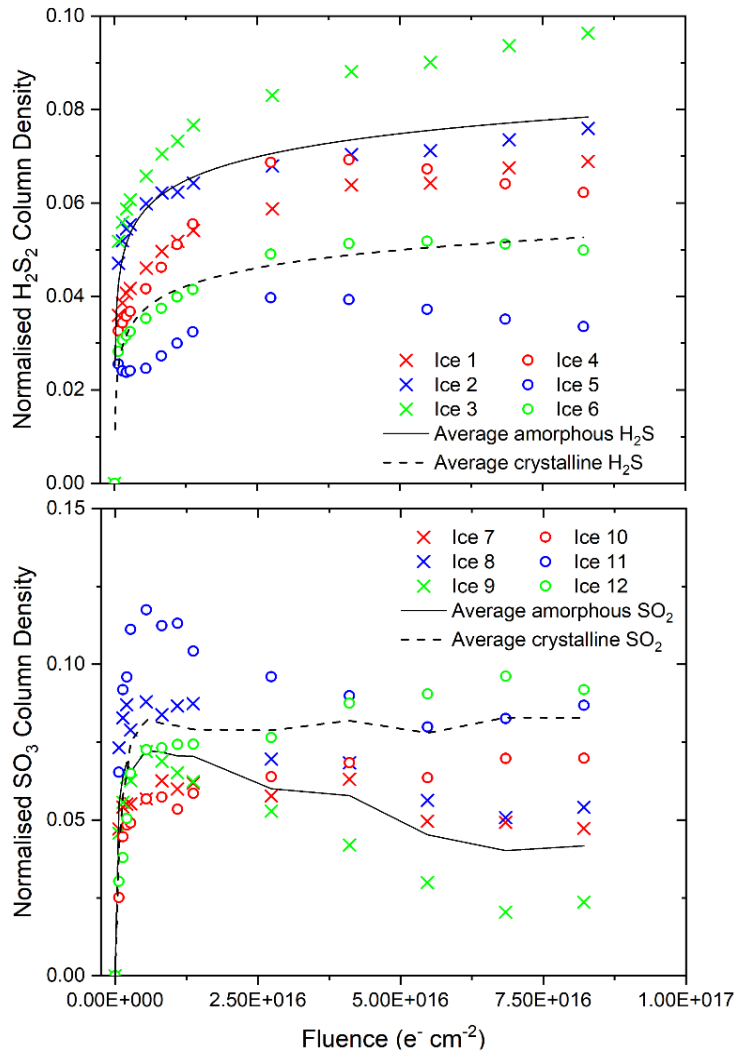

**Fig. 5:** *Above:* Column density of $H_2S_2$ from amorphous and crystalline $H_2S$ ices irradiated using 2 keV electrons at 20 K. *Below:* Column density of $SO_3$ from amorphous and crystalline $SO_2$ ices irradiated using 1.5 keV electrons at 20 K. Column densities have been normalised to the initially deposited column density of the parent molecular ice. Note that in the case of $H_2S_2$ the average trends are fitted by logarithmic functions while in the case of $SO_3$ the average trends are not fits and are plotted solely to guide the eye.

The sulphur budgets of each of the irradiated ices considered in this study are shown in Fig. 6. It is possible to note that a loss of sulphur is observed upon supplying an initial electron fluence of $6.9 \times 10^{14}$ electrons cm$^{-2}$ in all of the ices apart from the amorphous $SO_2$ ice, and that the quantity of unobserved sulphur as a fraction of that initially deposited continually grows during irradiation. In the case of the amorphous $SO_2$ ice, unaccounted for sulphur is only registered after a fluence of $2.7 \times 10^{16}$ electrons cm$^{-2}$ has been supplied, possibly due to the resistance of SO$_2$ to radiolytic dissociation as discussed earlier (de Souza Bonfim *et al.* 2017).

Although it is possible that electron irradiation resulted in the sputtering or desorption of the parent ice species, we consider this process to have likely been a relatively minor one. Previous work has shown, for example, that the reactive desorption of H$_2$S upon its formation as a result of the hydrogenation of HS on the surface of an interstellar ice analogue has a probability of 3% per hydrogenation event (Oba *et al.* 2018, Oba *et al.* 2019, Furuya *et al.* 2022), which is small from the perspective of an experimental study. Thus, if the electron-induced sputtering or desorption of sulphur-bearing molecules from the bulk ice is assumed to be negligible, then the fractions of unobserved sulphur shown in Fig. 6 represent the sulphur present in an infrared inactive form, such as atomic sulphur or, more likely, residues composed of sulphur allotropes (Gomis and Strazzulla 2008). Our data therefore suggest an important point with regards to the production of such residues from pure H$_2$S and SO$_2$ ices: it is apparent that the irradiation of amorphous ices results in a greater abundance of sulphur residues than does the irradiation of crystalline ices. It should be noted, however, that the conversion of observable molecular sulphur to unobservable residues is very efficient in each of the considered ices, with amorphous H$_2$S, crystalline H$_2$S, amorphous SO$_2$, and crystalline SO$_2$ ices respectively showing 41%, 25%, 44%, and 32% conversion of the initially deposited sulphur to residues at the end of irradiation (Fig. 6).

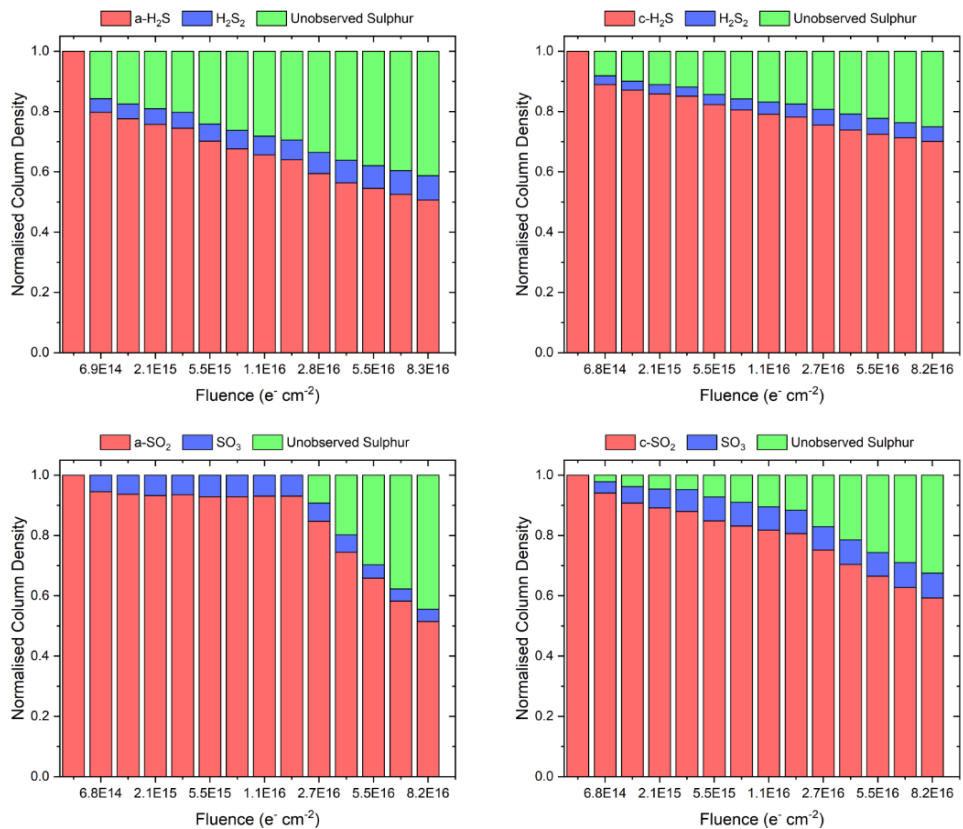

**Fig. 6:** Sulphur budgets of the electron irradiated amorphous and crystalline H$_2$S and SO$_2$ ices considered in this study. Uncertainties in the normalised abundance of the parent and primary product molecules are estimated to be within 3%. The quantity of unobserved sulphur represents an upper bound for the abundance of atomic or allotropic sulphur formed as a result of irradiation, since it is not known how many (if any) sulphur-containing species were sputtered or desorbed from the bulk ice. Note that the notations "a-" and "c-" used in the caption indicate whether the irradiated ice is amorphous or crystalline.

## 4   Implications for Interstellar and Solar System Chemistry

The results of this present study are directly applicable to the astrophysical chemistry of sulphur. In dense molecular clouds in the interstellar medium, there is a known paucity of observed sulphur relative to its expected cosmic abundance (Tieftrunk *et al.* 1994, Ruffle *et al.* 1999). Recent studies have suggested that this depletion may be mediated by the Coulomb-enhanced freeze-out of sulphur cations onto negatively charged dust grains, whereupon they polymerise to yield sulphur-bearing residues and chains (Cazaux *et al.* 2022). Modelling efforts have suggested different explanations as to the major forms of sulphur in interstellar space: Vidal *et al.* (2017) suggested that, depending upon the age of the dense cloud, the majority of the sulphur exists either as unobservable atoms in the gas phase or as $H_2S$ within icy grain mantles. Navarro-Almaida *et al.* (2020) also suggested the dominance of gas-phase sulphur atoms, whilst Laas and Caselli (2019) concluded that the majority of sulphur is found as organosulphur molecules within the icy grain mantles. Shingledecker *et al.* (2020) proposed that the major sulphur-bearing species in the condensed phase were sulphur allotropes along with $SO_2$ and OCS.

Nonetheless, it is expected that $H_2S$ will be present within icy grain mantles as a result of the hydrogenation of adsorbed sulphur atoms. Furthermore, $SO_2$ is suspected to be present in such ices on the basis of its tentative detection (Boogert *et al.* 1997). Our results demonstrate that the irradiation of these ices by galactic cosmic rays (for which we have used an energetic electron beam as a simulant) could further contribute to the presence of sulphur residues, chains, and atoms in the dense interstellar medium and, by extension, could also account for a portion of the depleted sulphur in such regions. Moreover, such processes are likely to be more efficient when the dense cloud is either fairly young (i.e., it is still in its pre-stellar stage) or in those regions of the cloud which are not in proximity to heat sources such as proto-stars since, under such conditions, the interstellar icy grain mantles would not undergo crystallisation or thermal segregation of their molecular constituents and would thus remain amorphous.

Our results are also applicable to outer Solar System chemistry, particularly in the cases of the Galilean moons of Jupiter and of comets. $SO_2$ is the dominant molecular component of the surface ices and exosphere of the innermost of the Galilean moons; Io (Douté *et al.* 2001), and has also been detected as a component of the icy surfaces of Europa, Ganymede, and Callisto (Noll *et al.* 1995, Noll *et al.* 1997, Domingue *et al.* 1998). Surface temperatures on Io undergo quotidian cycles between 90-130 K, thus allowing for cycles of sublimation and condensation of the surface $SO_2$ frosts to be maintained (Bagenal and Dols 2020). During the Ionian day, warmer temperatures cause the sublimation of much of the surface $SO_2$ ice, resulting in the formation of a tenuous exosphere. At night, however, lower temperatures drive the collapse of much of the exosphere and re-condensation of the $SO_2$ to surface ices.

Given that Io orbits within the giant Jovian magnetosphere, its surface is continually exposed to ionising radiation in the form of energetic ions and electrons. The flux of 0.1-52 keV electrons at the surface of Io was given by Frank and Paterson (1999) to be $3.1 \times 10^8$ electrons $cm^{-2}$ $s^{-1}$, meaning the fluence delivered in our experiments would be delivered to the Ionian surface within 8.5 years. The temperature conditions at the surface of Io would lead one to assume that $SO_2$ ice is naturally found in the crystalline phase, and that, therefore, the radiation-induced formation of $SO_3$ should be reasonably efficient (Fig. 5). However, our results also demonstrate that the prolonged irradiation of crystalline $SO_2$ ice at 20 K results in its amorphisation, reducing the comparative yield of $SO_3$ in favour of refractory residues of allotropic sulphur (Fig. 6). Such residues may contribute to the distinct colouration of Io (Carlson *et al.* 2007). It should be noted, however, that the extrapolation of radiation-induced amorphisation results acquired at low temperatures to higher ones may not be appropriate. For instance, although the amorphisation of crystalline $H_2O$ is known to occur efficiently as a result of its irradiation at 20 K, this process has never been reported at temperatures >70 K (Mifsud *et al.* 2022c). The efficiency of the radiation-induced crystalline $SO_2$ ice amorphisation process at various temperatures (including those relevant to the surface of Io) should therefore be tested in future experiments.

Finally, we note that our results are also applicable to the chemistry occurring within the icy nuclei of comets. The recent ESA *Rosetta* mission to comet 67P/Churyumov-Gerasimenko revealed the presence of a number of sulphur-bearing molecules within its icy nucleus, including $H_2S$, $SO_2$, SO, OCS, $CS_2$, and $S_2$ (Rubin *et al.* 2020). As the comet approaches perihelion in its orbit around the sun, thermally-induced crystallisation processes begin to out-compete space radiation-induced amorphisation. Correspondingly, the formation of allotropic sulphur residues *via* the irradiation of the $H_2S$ and $SO_2$ cometary ice components by the solar wind may decrease slightly in line with the results presented in Fig. 6.

## 5   Conclusions

In this experimental study, we have performed comparative and systematic electron irradiations of the amorphous and crystalline phases of $H_2S$ and $SO_2$ ices using 2 and 1.5 keV electrons, respectively. We have shown that, in the case of $H_2S$, the amorphous parent ice decays at a more rapid rate than does the crystalline one, in a manner that is similar to that previously reported for $CH_3OH$ (Mifsud *et al.* 2022b). This has been attributed to the presence of a more structured and extensive hydrogen bonding system in the crystalline phase compared to the amorphous phase, as well as the inherent lattice energy of the former, which require an additional energy input from the projectile electrons to be overcome before radiolytic chemistry may proceed. The formation of $H_2S_2$ as a product of the electron irradiation of $H_2S$ occurs to a greater extent in the amorphous phase than in the crystalline phase, in part due to the greater abundance of radiolytically generated HS radicals.

The irradiation of the $SO_2$ ice revealed unexpected results. In the amorphous ice, two regimes are apparent: a low-fluence regime in which the ice is possibly resistant to radiolytic decay (potentially due to the favourable reformation of excited $SO_2$ after the dissociation of ground-state $SO_2$) and a high-fluence regime in which a slow exponential-like decay trend is observed. This contrasts greatly with the crystalline ice, for which a rapid exponential decay is first observed in the low-fluence regime followed by a slower decay (which is slower than that of the amorphous phase) in the high-fluence regime. Interestingly, the formation of $SO_3$ as a result of the irradiation of the crystalline ice was always greater than during the irradiation of the amorphous ice, possibly due to the initial resistance of the amorphous ice to radiolytic decay and its subsequent preferential formation of infrared inactive sulphur allotropes and residues.

We suggest that our results are important not only in the context of further investigating the phase-dependent radiation chemistry of astrochemical ices, which has thus far been overlooked in the literature, but also in further understanding the chemistry of sulphur in extra-terrestrial environments. Our characterisation of this phase-dependent chemistry is directly applicable to understanding the sulphur chemistry on the surface of Io, as well as in the icy nuclei of comets. Moreover, our calculated sulphur budgets for each of the irradiation processes considered in this study (which reveal the seemingly efficient formation of infrared inactive sulphur allotropes and residues) may aid in further constraining the exact molecular forms of sulphur in interstellar icy grain mantles and cometary ices. Finally, we conclude by noting that our experimental results further demonstrate the importance of incorporating ice phase as a factor when designing more complete 'systems astrochemistry' investigations (Mason *et al.* 2021).

## Author Contributions

The experiment was designed by Duncan V. Mifsud and Perry A. Hailey and carried out by Duncan V. Mifsud, Péter Herczku, Sándor T.S. Kovács, Zoltán Juhász, and Béla Sulik. Data analysis was performed by Duncan V. Mifsud, who also wrote the manuscript. All authors were responsible for results interpretation and improvements to the manuscript.

## Acknowledgements

The authors gratefully acknowledge funding from the Europlanet 2024 RI which has been funded by the European Union Horizon 2020 Research Innovation Programme under grant agreement No. 871149. The main components of the experimental apparatus were purchased using funding obtained from the Royal Society through grants UF130409, RGF/EA/180306, and URF/R/191018. Recent developments of the experimental set-up were supported in part by the Eötvös Loránd Research Network through grants ELKH IF-2/2019 and ELKH IF-5/2020. We also acknowledge support from the National Research, Development, and Innovation Fund of Hungary through grant No. K128621. Duncan V. Mifsud is the grateful recipient of a University of Kent Vice-Chancellor's Research Scholarship. The research of Zuzana Kaňuchová is supported by the Slovak Grant Agency for Science (grant No. 2/0059/22) and the Slovak Research and Development Agency (contract No. APVV-19-0072). Sergio Ioppolo acknowledges the Royal Society for financial support. The authors would also like to thank Béla Paripás (University of Miskolc, Hungary) for his continued support and assistance.